\pdfoutput=1
\documentclass{article}
\usepackage[utf8]{inputenc}

\newenvironment{eqaed}
    {
    \begin{equation}
        \begin{aligned}
    }
    {
        \end{aligned}
    \end{equation}\ignorespacesafterend
    }

\usepackage{amsmath,amsthm,amssymb}
\usepackage[dvipsnames]{xcolor}
\usepackage{physics}
\usepackage{tensor}
\usepackage{graphicx}
\usepackage{float}
\usepackage{xcolor, colortbl}
\usepackage{hyperref}
\usepackage{authblk}
\usepackage{mathtools}
\usepackage{geometry}
\definecolor{light-gray}{gray}{0.9}

\usepackage{setspace}
\usepackage{xparse}
\usepackage{esvect} 
\usepackage{slashed} 
\usepackage{dsfont} 
\usepackage[dvipsnames]{xcolor} 
\usepackage{empheq}
\usepackage{makecell}
\usepackage{subcaption}
\usepackage{float}
\usepackage{hyperref}
\hypersetup{
    colorlinks=true,
    linkcolor=blue,
    filecolor=magenta,      
    urlcolor=cyan,
}
\usepackage{import}
\usepackage{makeidx}
\title{\sc{Analysis of Black Hole Solutions in Parabolic Class Using Neural Networks}}

\author[1]{ Ehsan Hatefi\footnote{ehsan.hatefi@uah.es, ehsanhatefi@gmail.com}}
\author[2]{ Armin Hatefi\footnote{ahatefi@mun.ca }}
\author[1]{Roberto J. L\'opez-Sastre\footnote{robertoj.lopez@uah.es}}
\affil[1]{University of Alcal\'a, Department of Signal Theory and Communications, Research group GRAM, Alcal\'a de Henares, Spain.}
\affil[2]{Department of Mathematics and Statistics, Memorial University of Newfoundland, St John’s, NL, Canada.}

\begin{document}

\maketitle

\vspace{-0.7cm}

\begin{abstract}
In this paper, we introduce a numerical method based on Artificial Neural Networks (ANNs) for the analysis of black hole solutions to the Einstein-axion-dilaton system in a high dimensional parabolic class. Leveraging a profile root-finding technique based on General Relativity we describe an ANN solver to directly tackle the system of ordinary differential equations. 
Through our extensive numerical analysis, we demonstrate, for the first time, that there is no self-similar critical solution for  the parabolic class in the high dimensions of space-time.
Specifically, we develop 95\% ANN-based confidence intervals for all the solutions in their domains. At the 95\% confidence level, our ANN estimators confirm that there is no black hole solution in higher dimensions, hence the gravitational collapse does not occur. 
Results provide some doubts about the universality of the Choptuik phenomena. Therefore, we conclude that the fastest-growing mode of the perturbations that determine the critical exponent does not exist for the parabolic class in the high dimensions.
\end{abstract}

\newpage

\section{Introduction}
 
It is well known that three parameters play key roles in describing black holes. These parameters include the mass, the angular momentum and the charge of the black holes. According to \cite{Chop}, there still may be another parameter that may be relevant to black holes and their solutions. Christodoulou \cite{Christodolou} studied the existence of the spherically symmetric collapse solutions to the Einstein scalar field. 
Choptuik \cite{Chop} then proposed a distinguished property of the collapse solutions illustrating the discrete self-similarity of the real scalar field.
Indeed, one can argue that various gravitational collapse solutions represent space-time self-similarity; hence dilations happen.

Note that all the critical solutions illustrate the scaling law. The initial condition of the scalar field can be characterized by the parameter $p$ relating to the field amplitude. Let $p=p_\text{crit}$ represent the critical solution. The black hole is then formed when $p$ takes on a value larger than $p_\text{crit}$. When $p$ becomes bigger than the critical value, the following scaling law determines the mass of the black hole or the Schwarzschild radius in $d \geq 4$ by
\begin{equation}
 r_S(p) \propto (p-p_\text{crit})^\gamma \,, \quad M_\text{bh}(p) \sim (p-p_\text{crit})^{(D-3)\gamma} \,,
\end{equation}
where $\gamma\simeq 0.37$ for four dimensions and for a real scalar field \cite{Chop,Hamade:1995ce,KHA,AlvarezGaume:2006dw}.

Several approaches investigated the numerical simulations for studying the spherically symmetry of the scalar field for some matter content \cite{Birukou:2002kk,Sorkin:2005vz,Bland:2005vu,HirschmannEardley,Rocha:2018lmv}. Moreover, in \cite{AlvarezGaume:2008qs,evanscoleman,KHA,MA} we find studies that evaluated the gravitational collapse solutions of the perfect fluid. The critical exponent $\gamma \simeq 0.36$ was first found in \cite{evanscoleman}. Strominger and Thorlacius \cite{Strominger:1993tt} discussed that $\gamma$ may have a universal value for all fields that can be coupled to gravity in four dimensions.

Maison \cite{MA}  has already shown the matter-dependence of the mass-scaling exponent.  
Hatefi and Kuntz \cite{Hatefi:2020gis} recently showed that the
critical exponent depends on the space-time dimension and also they confirmed that the 
critical exponent depends an the matter content and the
 ans\"atze of the various solutions of self-similar collapse.
These results thus provide doubts concerning the universality of the Choptuik exponent. It should be mentioned that one can derive the exponent based on various perturbations of self-similar solutions \cite{KHA,MA,Hirschmann:1995qx}. In fact, one can start perturbing field $h$ as 
\begin{equation}
    h = h_0 + \varepsilon \, h_{-\kappa} \;,
\end{equation}
where $h_{-\kappa}$ includes the scaling $-\kappa \in \mathbb{C}$ related to different modes. Note that the minus sign indicates a growing mode near the black-hole formation time as $t \rightarrow 0$. When one requires to find the most relevant mode $\kappa^*$ , corresponding to the highest value of $\Re(\kappa)$, $\gamma$ can be obtained from \cite{KHA,MA,Hirschmann:1995qx} by
\begin{equation}
    \gamma = \frac{1}{\Re \kappa^*}\, .
\end{equation}

Abrahams and Evans \cite{AE} explored numerically the axisymmetric gravitational collapse. Alvarez-Gaume et al. \cite{AlvarezGaume:2008fx} found the other critical solutions as well as the shock waves. Hirschmann and Eardley \cite{Hirschmann_1997} derived the original results for the axion-dilaton in four dimensions and concluded that $\gamma \simeq 0.2641$.

There is a relationship between the axion-dilaton system and the so-called formation of black holes, particularly in analyzing their holographic descriptions in diverse dimensions, see \cite{AlvarezGaume:2006dw}. The axion-dilaton system is relevant to the gauge/gravity correspondence \cite{Maldacena:1997re}, relating the critical exponent and Choptuik exponent, the imaginary part of quasinormal modes and the dual conformal field theory \cite{Birmingham:2001hc}. This axion-dilaton system has found applications in various areas, including the physics of black holes \cite{Hatefi:2012bp,Ghodsi_2010} as well as the S-duality transformation in this particular class \cite{Hamade:1995jx}. Another application of the axion-dilaton system appears in cases where one is interested in the collapse of spaces that can asymptotically tend to $AdS_5 \times S^5$. In these cases, the matter content is given by the axion-dilaton system and the self-dual 5-form field. 
 
Alvarez-Gaume and Hatefi \cite{AlvarezGaume:2011rk} originally investigated the gravitational collapse of the axion-dilaton system in the elliptic case. They computed the critical collapse of the system by analyzing the hyperbolic ansatz \cite{hatefialvarez1307}. Subsequently, the continuous self-similar solutions were calculated in \cite{ours} in various dimensions. In a recent paper, Antonelli and Hatefi \cite{Antonelli:2019dqv} derived the perturbations and reproduced the correct value of $\gamma \sim 0.2641$ in four dimensions \cite{Hirschmann_1997}. Later on, in \cite{Hatefi:2020gis} a computation of the perturbation equations and the extraction of other critical exponents was studied. 

Recently, in \cite{Hatefi:2022shc,Hatefi:2021xwh} Fourier-based regression models and nonlinear statistical spline smothers are explored for estimating the gravitational collapse functions of the axion-dilaton system under 4-dimensional elliptic and hyperbolic spaces. While these statistical methods perform well enough in estimating the nonlinear pattern of the critical collapse functions in four dimensions case, the Fourier-based estimators may change significantly in the case of slight changes in the estimates of the regression parameters. Another challenge associated with the spline regression methods is that the methods require realizations of the underlying population. In other words, one has to solve the equations of motion iteratively to observe the response from the critical collapse functions on their entire domain. This may be challenging when finding a solution to the equations of motion results in numerical instability, or there may not be a clear solution to distinguish the domains of the critical collapse functions.

Previous works focus the analysis of the solutions for the axion-dilaton system in low dimensional space-time. Our first contribution is that in this work we extend the study of these solutions, under the parabolic class, to higher dimensional space-time, ranging from four to nine.
Given the recent advances in Artificial Neural Networks (ANNs) for solving differential equations \cite{Chen:2020,DeepXDE,pydens_2019}, we propose, for the first time, an ANN-based solver to estimate the self-similar gravitational collapse solutions of Einstein axion-dilaton system under parabolic class. Leveraging a profile root-finding technique based on General Relativity, our approach is able to directly tackle the system of ordinary differential equations associated to the axion-dilaton system.
Unlike previous works in the literature on critical solutions in parabolic class, we develop a stochastic approach to present the universal estimator, modeled by the ANN, of the critical collapse functions to deal with the sampling variability involved in neural networks for black hole solutions in parabolic class. In other words, we construct 95\% empirical intervals for the estimators of all the critical collapse functions in parabolic case. Through extensive numerical studies, at a 95\% confidence level, we show that there is no black hole solution in higher dimensions for the parabolic class using our ANN-based solver.

This paper is organized as follows. Section \ref{sec:equ_motion} describes the self-similar axion-dilaton system in $d$ dimensions and the equations of motion for the third conjugacy class of SL(2,R) transformation. Section \ref{sec:NNs} explains our approach by using ANNs to analyze the solutions to the the self-similar axion-dilaton system. In Section \ref{sec:num_studies}, we investigate the performance of the ANN-based method in estimating the critical collapse functions under parabolic class for various space-time high dimensions. Section \ref{sec:sum}, finally, presents a summary and concluding remarks.

 \section{Equations of Motion of Black Holes} \label{sec:equ_motion}
In this section, we discuss the self-similar axion-dilaton system and then develop the equations of motion for the third conjugacy class of SL(2,R) transformation.
\subsection{The Relevant System in Parabolic Class}\label{subsec:unperturbed}
 The effective action of the Einstein axion-dilaton system that can be coupled to $d$ dimensional gravity \cite{Sen:1994fa,Schwarz:1994xn} is given by
 \begin{equation}
S = \int d^d x \sqrt{-g} \left( R - \frac{1}{2} \frac{ \partial_a \tau
\partial^a \bar{\tau}}{(\Im\tau)^2} \right) \;.
\label{eaction}
\end{equation}
The axion-dilaton can be combined by defining $\tau \equiv a + i e^{-\phi}$. The SL(2,R) symmetry of action in Eq. \eqref{eaction} is shown by
\begin{equation} \label{eq:sltr}
     \tau \rightarrow M \tau \equiv \frac{a \tau +b}{c \tau + d}, \quad\quad ad-bc=1 \; ,
\end{equation}
where $a,b,c$ and $d$ are real parameters.

When the quantum effects are taken into account, the SL(2,R) symmetry will be related to $\mathrm{SL}(2,\mathbb{Z})$ and this duality also holds for the case of the non-perturbative symmetry \cite{gsw,JOE,Font:1990gx}. One can obtain the equations of motion by taking the variations from the metric and the $\tau$ field, respectively. Hence the equations of motion are given by
\begin{equation}
\label{eq:efes}
R_{ab} = \frac{1}{4 (\Im\tau)^2} ( \partial_a \tau \partial_b
\bar{\tau} + \partial_a \bar{\tau} \partial_b \tau)\;,
\end{equation}
\begin{equation}\label{eq:taueom}
\nabla^a \nabla_a \tau +\frac{ i \nabla^a \tau \nabla_a \tau }{
\Im\tau} = 0 \,.
\end{equation}

The metric in $d$ dimensions for the spherical symmetry can be written as
\begin{equation}
    ds^2 = (1+u(t,r)) (-b(t,r)^2 dt^2 + dr^2) + r^2 d\Omega_{d-2}^2 \,,
\end{equation}
where $\tau$ field represents a function of both coordinates of $t$ and $r$, that is $ \tau = \tau(t,r)$, and $d\Omega_{d-2}^2$ denotes the angular part for the metric.

If we use the scale-invariant $(t,r)\rightarrow ( \Lambda t,\Lambda r)$, then  $ ds^2$ must scale as $ds^2 \rightarrow \Lambda^2 ds^2$. Thus all the functions in the metric should be scale-invariant, namely $u(t,r) = u(z)$, $b(t,r) = b(z)$, $z \equiv -r/t$. 
The effective action in Eq. \eqref{eaction} is invariant under the SL(2,R) transformation \eqref{eq:sltr}. This implies that
$\tau$ should also be invariant up to an SL(2,R) transformation as
\begin{equation}\label{tau12}
    \tau(\Lambda t, \Lambda r) = M(\Lambda) \tau(t,r) \;.
\end{equation}
When the system of $(g,\tau)$ enjoys the above invariant properties, then the system is called a continuous self-similar (CSS) solution. Note that the different assumptions can be related to different classes of $\eval{\dv{M}{\Lambda}}_{\Lambda=1}$~\cite{ours}, where $\tau(t,r)$ can take the elliptic, hyperbolic and parabolic assumptions. 
From \eqref{tau12}, the general form of the ansatz for the elliptic, hyperbolic and parabolic class is given by
\begin{equation}\label{eq:tauansatz}
    \tau(t,r) = \begin{dcases}
                i \frac{1-(-t)^{i\omega}f(z)}{1+(-t)^{i\omega} f(z)}\,, & \quad \text{elliptic}\\[5pt]
                    \frac{1-(-t)^{\omega}f(z)}{1+(-t)^{\omega} f(z)}\,,& \quad \text{hyperbolic}\\[5pt]
                    f(z) + \omega \log(-t)\,, & \quad \text{parabolic}
                \end{dcases}
\end{equation}
where $\omega$ is an unknown real parameter for all the classes. Also function $f(z)$ satisfies the condition $\abs{f(z)} < 1$ for the elliptic case; however, function $f(z)$ satisfies the condition $\Im f(z)>0$ for hyperbolic and parabolic classes. 

It is worth mentioning that in Appendix \ref{sec:appendix} we show that the proposed ANNs-based approach can confirm that there exists a solution in the four-dimensional elliptic case which is compatible with findings of \cite{ours}. Here, we just study the parabolic self-similar solutions in detail. From the equations of motion \eqref{eq:efes} and \eqref{eq:taueom}, it can  be shown that the following symmetry is present. In other words, if
\begin{equation}
f(z) \rightarrow f(z) + a \,,
\end{equation}
then all the equations of motion remain invariant.

Applying CSS ans\"atze \eqref{eq:tauansatz} to equations of motion \eqref{eq:efes} and \eqref{eq:taueom}, one can develop the ordinary differential equations for $u(z)$, $b(z)$, $f(z)$. If we take the spherical symmetry, $u(z)$ can be eliminated in terms of $b(z)$ and $f(z)$, employing
\begin{equation}\label{eq:u0explicit}
u(z) = \frac{z b'(z)}{(q-1) b(z)}\,.
\end{equation}

Note that the first derivative of $u(z)$ can also be removed from all equations of motion as
\begin{equation}\label{eq:u0pexplicit}
   \frac{qu'(z)}{2(1+u(z))} = 
   \frac{w \bar f(z) f'(z)+w  f(z) \bar f'(z) - 2z  \bar f'(z) f'(z)}{(f(z)-\bar f(z))^2}\,.
\end{equation}

Finally, all the ordinary differential equations (ODEs) are given by
\begin{align}\label{eq:unperturbedbp}
    b'(z) & = B(b(z),f(z),f'(z))\,, \\
    f''(z) & = F(b(z),f(z),f'(z)).
    \label{eq:unperturbedfpp}
\end{align}

From \eqref{eq:efes} and \eqref{eq:taueom}, the equations of motion for self-similar solutions for the parabolic case in $d$ dimensions $(d=4,..,10)$ are given by the following system
\begin{eqnarray}
0 & = & b'-{ 2z(b^2 - z^2) \over (d-2)b (f -\bar f)^2} f' \bar{f}' + {
2 \omega (b^2 - z^2) \over (d-2)b (f -\bar f)^2} ( \bar{f}'+ f')
+ {2\omega^2 z  \over (d-2)b (f -\bar f)^2}, \nonumber\\
0 & = & -f''
     - {2z (b^2 + z^2) \over (d-2)b^2 (f -\bar f)^2} f'^2 \bar{f}'
     + {2 \over (f -\bar f)} \left(1 
       + { \omega (b^2 + z^2) \over (d-2) b^2 (f -\bar f)} \right) f'^2 \nonumber \\&&
     + {2 \omega (b^2 + 2 z^2) \over (d-2)b^2 (f -\bar f)^2}  f'
\bar{f}' 
  + {2 \over z} \left(-\frac{(z^2-\frac{(d-2)b^2}{2})}{(z^2-b^2)} + {2\omega z^2  \over (b^2 - z^2)
(f -\bar f)}\right.\nonumber \\&& 
+ \left.{2\omega^2 z^4  \over (d-2)b^2 (b^2 - z^2) 
(f -\bar f)^2}\right) f'- {2\omega^2 z \over (d-2)b^2 (f -\bar f)^2} 
\bar{f}' + \nonumber \\&&
{2 \omega \over (b^2 - z^2)} \left(-\frac{1}{2} - { \omega
\over (f -\bar f)}\right.
- \left.{\omega^2 z^2  \over (d-2)b^2 (f -\bar f)^2}
\right).
\label{1fzeom321}
\end{eqnarray}

These equations are invariant under some shifts of $f(z)$ by a real number. Let $f(z)=u(z)+i v(z),\bar f(z)=u(z)-i v(z)$ where $u(z), v(z)$ are real functions. Solving the system of equations \eqref{1fzeom321} in Cartesian coordinates, the equations of motion, for $d$ dimensions  $(d = 4, .., 10)$ can be written as follows:

\begin{equation}\label{eq:b0}
-2(d-2)b'bv^2+ w^2z+(z^2-b^2)(-2wu'+zu'^2+zv'^2)=0\,.
\end{equation}

The other equation for $f''$ is given by

\begin{eqaed} \label{eq:f0''}
2(d-2)b^2v^2(u''+iv'') z(z^2-b^2)+w^3 z^3-2(d-2)v^2wzb^2-3w^2z^4u'+4(d-2)v^2z^2b^2u'\\+w^2z^2b^2u'-2(d-2)^2v^2b^4u'+3wz^5u'^2-wz^3b^2u'^2-2wzb^4u'^2\\
-z^6u'^3 +z^2b^4u'^3+4(d-2)vwz^2b^2v'-4(d-2)vz^3b^2u'v'\\
+4(d-2)vzb^4u'v'+wz^5v'^2-wz^3b^2v'^2-z^6u'v'^2 +z^2b^4u'v'^2\\
+i\bigg(2(d-2)vw^2zb^2-4(d-2)vwz^2b^2u'+2(d-2)vz^3b^2u'^2-2(d-2)vzb^4u'^2-w^2z^4v'\\+4(d-2)v^2z^2b^2v'-w^2z^2b^2v'-2(d-2)^2v^2b^4v'+2wz^5u'v'-2wzb^4u'v'\\
-z^6u'^2v'+z^2b^4u'^2v'-2(d-2)vz^3b^2v'^2+2(d-2)vzb^4v'^2-z^6v'^3+z^2b^4v'^3
    \bigg)=0 \,.
\end{eqaed}

The regularity of $\tau$ and residual symmetries in the equations of motions \eqref{1fzeom321} entails the initial boundary conditions as
\begin{align}\label{boundry_bf} 
\left\{\begin{array}{lc} 
b(0) = 1, &  f'(0) =0\,,  \\
f(0) = i \Im f(0), &  \Im f(0)= x_0\,, 
\end{array}\right.
\end{align}
where $x_0$ is a real parameter and $\Im f(0)>0$. 

Hence, the problem is reduced to determine two real parameters, including $\omega$ and $f(0)$. The equations of motion in \eqref{1fzeom321} have five singularities (see \cite{AlvarezGaume:2011rk}) located at $z = \pm 0$, $z = \infty$ and $z = z_\pm$. The last two singularities are expressed by $b(z_\pm) = \pm z_\pm$. They are related to the homothetic horizon where $z=z_+$ is just a mere coordinate singularity \cite{Hirschmann_1997,AlvarezGaume:2011rk}. Thus, $\tau$ must be regular across it, representing that $f''(z)$ must be finite as $z\rightarrow z_+$. Therefore, the vanishing of the divergent part of $f''(z)$ generates a complex-valued constraint at $z_+$, denoted by  $G(b(z_+), f(z_+), f'(z_+)) = 0$. The explicit form of the $G$ function for the parabolic case is then given by 
\begin{eqaed}\label{eq:Gparabolic2}
G(f(z_+),f'(z_+))   = & \, -\frac{1}{f(z_+)}\Bigg(-2 (d-2) \omega \bar{f}(z_+) \left(\omega-2 z_+ f'(z_+)\right)\\& +(d-2) \bar{f}(z_+)^2 \left((d-4) z_+ f'(z_+)+\omega\right)\\&-2 (d-2) f(z_+) \left(\bar{f}(z_+) \left((d-4) z_+ f'(z_+)+\omega\right)-\omega \left(\omega-2 z_+ f'(z_+)\right)\right)\\&+(d-2) f(z_+)^2 \left((d-4) z_+ f'(z_+)+\omega\right)+2 \omega^2 \left(\omega-2 z_+ f'(z_+)\right)\Bigg)\,.
\end{eqaed}

Owing to the vanishing of the real and imaginary parts of $G$, one can rewrite the equations in \eqref{1fzeom321} as a system of equations based on two unknown parameters $(\omega,\Im f(0))$. Once the system is obtained, using $f(0)=i x_0$ and the boundary condition at $z=0$, one can find the discrete solutions by integrating numerically the equations of motion \eqref{1fzeom321}. For instance, the discrete solution in four dimensions for the elliptic case was first found in \cite{Eardley:1995ns} and then was confirmed in \cite{AlvarezGaume:2011rk, ours}.

\subsection{Scaling Symmetry}
 One can show that there is an extra symmetry for the parabolic case from the ansatz as
\begin{equation}
\tau(t,r) = f(z)+\omega \log(-t) \,.
\end{equation}

If we scale $\omega$ and $f(z)$ by a constant $L$, then $\tau$ also scales by $L$. In other words, if we have
\begin{equation}\label{parabolic_rescaling}
    \omega \rightarrow L \omega\,,\quad f(z) \rightarrow L f(z)\,,\quad L \in \mathbb{R}_+\,,
\end{equation}
then $\tau$ also transforms as $\tau\rightarrow L \tau$.  Hence, if $(\omega,\Im f(0))$ is a solution, then  $(L \omega,L\Im f(0))$ leads to the same solution. 

Note that $G$ equation \eqref{eq:Gparabolic2} scales homogeneously as $L^2$ under $(\omega,f)\rightarrow (L\omega, L f)$, because all equations of motion and the constraint $G(\omega,\Im f(0))$ are invariant under this new scaling. According to this degeneracy, the only real unknown parameter becomes the ratio $\omega/\Im f(0)$, and the two-dimensional constraint should be solved for it. 

Due to the fact that $G(\omega,Im f(0))$ only depends on $\omega/ \Im f(0)$, one can apply a profile root-finding method to determine the solutions to the equation of $G$. To do that, we first set $\Im f(0)=1$ and treated equation \eqref{eq:Gparabolic2}  as a function of $\omega$, and next found the root of $G$ on $\omega$ coordinate. If there exists a root, then it generates a continuous ray of solutions $(\omega, \Im f(0)) = (K\omega^*, K)$. Different solutions in the same ray must be related to an SL(2,R) transformation where they are not physically distinguishable. When $\tau (t, r) = \tau(z)$, we can show that this trivial solution can be obtained from any of the previous solutions by $\omega \rightarrow 0$, using the initial boundary conditions  $b(0) = 1$, $\tau'(0)=0$. Therefore, $\tau(0)=\tau_0$ should be an arbitrary constant. These are, in fact, the initial conditions that reproduce flat space-time with a constant axion-dilaton of $\tau_0$.
\begin{figure}
    \centering
    \includegraphics[width=6in]{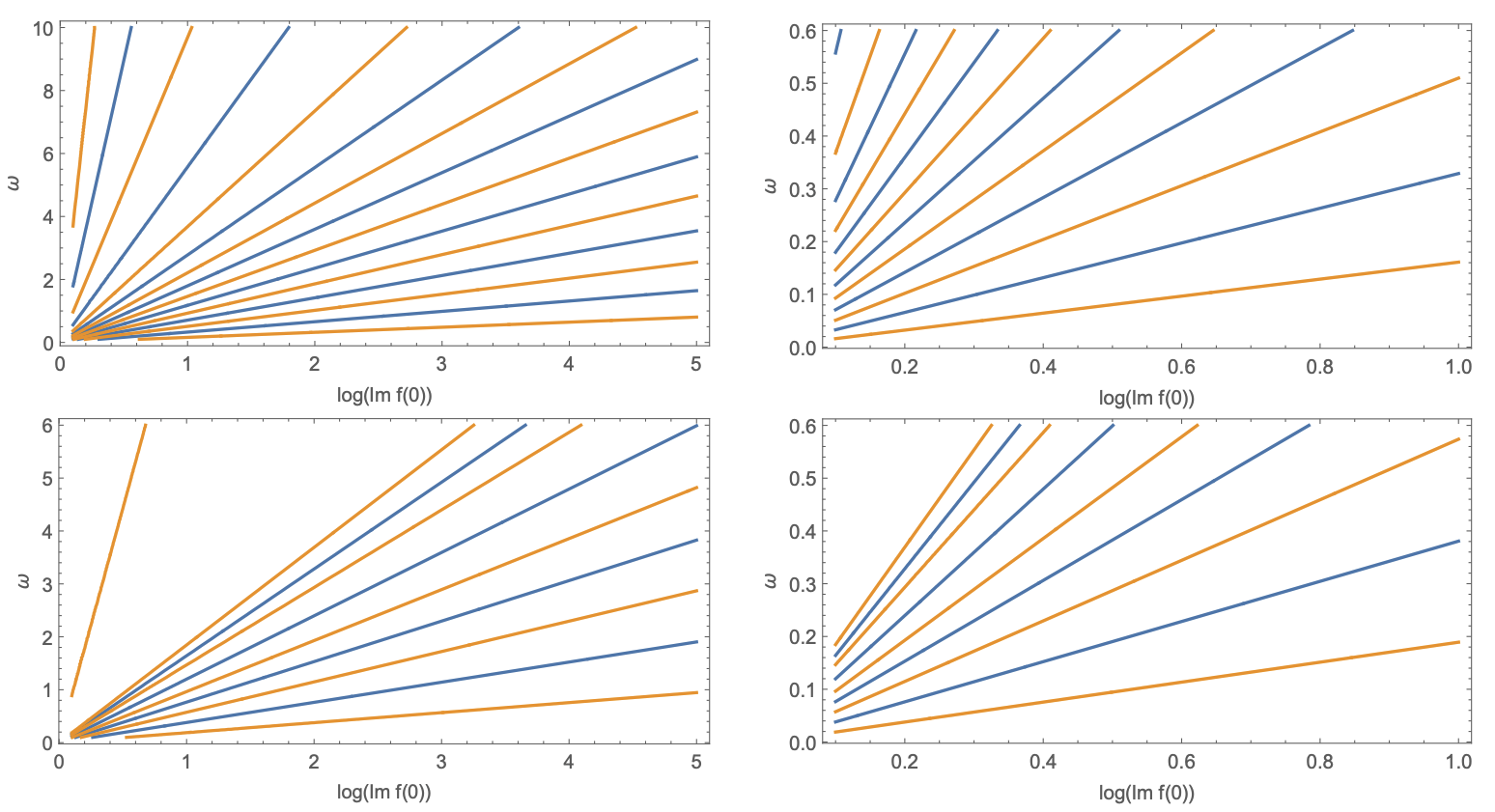}
    \caption{The contours of the real and imaginary parts of $G(\omega, \Im f(0))$ in four-dimensional (top panels) and five-dimensional (bottom panels) parabolic class. The blue and orange curves represent the real and imaginary functions, respectively.}
    \label{math_45d}
\end{figure}

The top panel of Figure \ref{math_45d} shows two-dimensional contours of the zeroes of the real part (with blue colour) and imaginary  part (with orange colour) of $G(\omega, \Im f(0))$, as functions of $(\omega, \log \Im f(0))$ in four-dimensional parabolic space. We observe that the curves corresponding to the real and imaginary parts of the $G$ function do not intersect in the entire region.  
The top panel of Figure \ref{math_45d} demonstrates the degeneracy associated with the extra scaling invariance \eqref{parabolic_rescaling} in four dimensions, as described earlier. 
According to the wide region of  $(\omega, \log \Im f(0))$ in the top-left panel of Figure  \ref{math_45d}, one may argue that there may exist a solution for the small values of  $\omega$ and $\log \Im f(0)$. To elaborate on this argument, we show the contours of the real and imaginary parts of $G(\omega, \Im f(0))$ but now on smaller values of $\omega$ and $\log \Im f(0))$ in the top-right panel of Figure \ref{math_45d}. From the later plot,  there is clearly no intersection between the real and imaginary parts of the $G$ function. This verifies that no solution exists to equations in \eqref{1fzeom321} in the four-dimensional parabolic space for $\omega > 0$.  

\subsection{Solutions in Higher Dimensions}
\label{sec:solutions_in_hd}
In this subsection, we investigate the solutions to equations of motion in \eqref{1fzeom321} for parabolic space in various space-time dimensions ranging from five to nine. Similarly to the four dimensions, one can use the constraints on vanishing the real and imaginary parts of $G(\omega, \Im f(0))$ and re-write equation \eqref{1fzeom321} as a function of two unknown real parameters $\omega$ and $\Im f(0)$. Using an appropriate starting point for $\Im f(0)$, one can apply the profiling method and find the root of  $G$ function \eqref{eq:Gparabolic2}.

\begin{figure}
    \centering
    \includegraphics[width=6in]{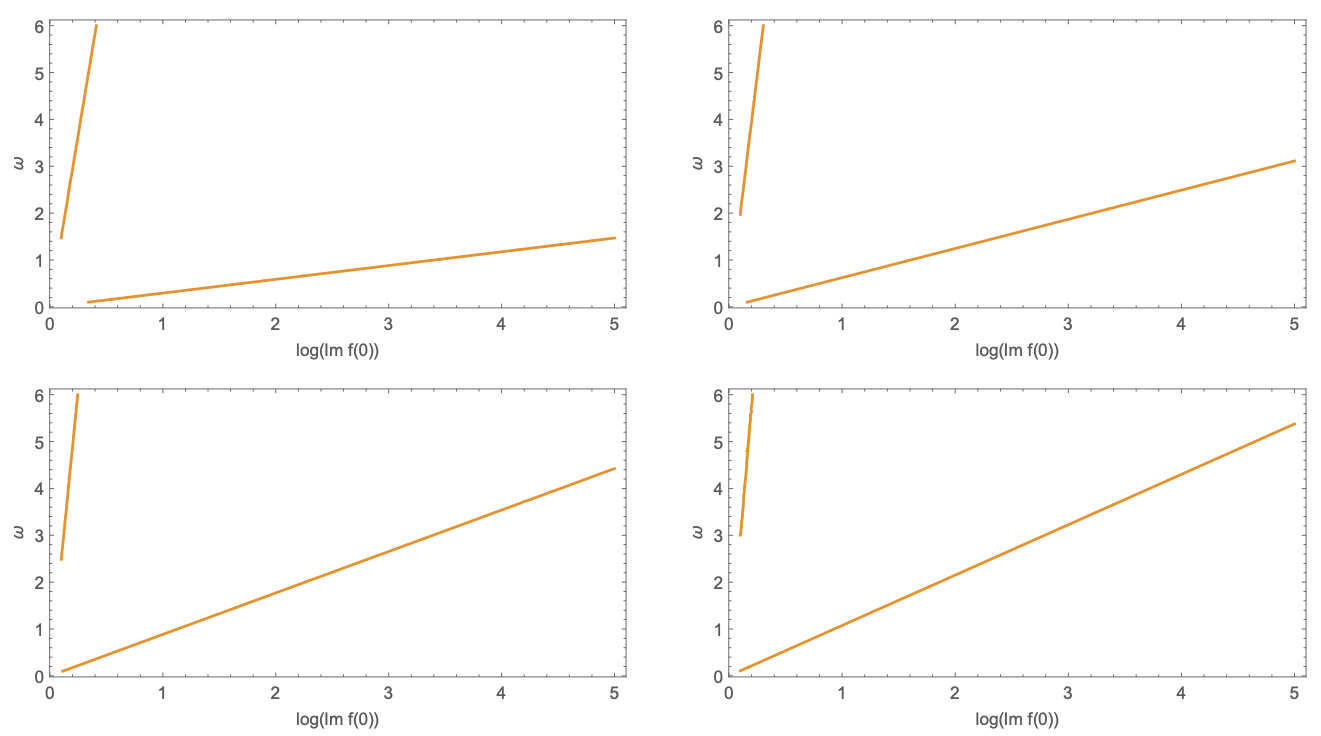}
    \caption{The contours of the real and imaginary parts of  $G(\omega, \Im f(0))$ in parabolic case for various dimensions including $d=6$ (top left), $d=7$ (top right), $d=8$ (bottom left) and $d=9$ (bottom right). 
    }
    \label{math_d6789}
\end{figure}

Figure \ref{math_45d} (bottom panel) and Figure \ref{math_d6789} demonstrate the contours using the zeroes of the real and imaginary parts of $G(\omega, \Im f(0))$ function \eqref{eq:Gparabolic2} for different dimensions ranging from five to nine. In each plot, blue and orange curves indicate the real and imaginary parts of the $G(\omega, \Im f(0))$ function, respectively. From Figures \ref{math_45d} and \ref{math_d6789} demonstrate that the two parts do not intersect throughout the entire domain of the parameters $(\omega, \log \Im f(0))$. This concludes that no parabolic black hole solution exists even in higher space-time dimensions $d \in \{5,\ldots,9\}$. This leads to the fact that gravitational collapse will not take place in high-dimensional parabolic cases and it leads to the  Minkowski flat space-time. 

Note that, for the five-dimensional parabolic case, one can search for the zeroes of the real and imaginary parts of $G(\omega,1)$ and may find a spurious solution ray. This solution ray may stem from the numerical instability of the equations of motion for a very small value of $\omega$. For further remarks in this case, readers are referred to \cite{Hatefi:2020jdr}. While the numerical accuracy is insufficient to assess the solution with certainty in five dimensions, using the root-finding method, one may identify a very small value for $|G|$ and a value of $\omega$ as
\begin{eqaed}
   \abs{G} \sim 0.006,\quad   \omega \sim 1.65 \label{math12} \; .
\end{eqaed}

It is evident from Figure \ref{math_d6789} that the higher the space-time grows, the more divergence is observed between the real and imaginary parts of the $G(\omega, \Im f(0))$ function. In other words, in the higher dimensions, the real part of the $G$ function diverges so quickly and disappears in the entire domain, that it is more likely that there is no interaction between the real and imaginary curves. Therefore, it confirms that no parabolic black hole solution is expected in higher dimensions.

\section{Search for Solutions Using Neural Networks}
\label{sec:NNs}

In this section, we describe an ANN-based solver for the equations of motion in \eqref{eq:b0} and \eqref{eq:f0''}. We intend to provide an alternative approach to the method described, in order to shed more light on the problem, confirming that no black hole solution exists in higher dimensional parabolic cases.

In the simplest form, an artificial neural network consists in a multi-layer perceptron \cite{Goodfellow:Deep}. The neural network utilizes linear and nonlinear transformations to dispatch data from one layer to another, ultimately from the input layer to the output layer. Many variations of neural networks have been developed, such as fully connected neural networks, convolutional neural networks, and recursive neural networks, to name a few \cite{Goodfellow:Deep,Ramsundar}. Artificial neural networks have found applications in almost all scientific fields ranging from bioinformatics \cite{Min} and material sciences \cite{Choudhary:Recent}, to natural language processing \cite{Collobert:unified}. In this work, we focus on the fully connected neural networks which are sufficient to deal with the non-linearity of the Einstein-axion-dilaton system in the high-dimensional parabolic classes. 

Let $\mathcal{N}^{L}({\bf x},t,{\bf \phi})$ represent a deep neural network with $L$ layers mapping $\mathbb{R}^{d_{\text{in}}} \rightarrow \mathbb{R}^{d_{\text{out}}}$. $d_{\text{in}}$ and $d_{\text{out}}$ encode the dimensionality of the input and output of the network. 
Let $n_l$ denote the number of neurons for the $l$-th layer of the network, for $l=1,\ldots, L-1$. $n_0=d_{\text{in}}$ and $n_L=d_{\text{out}}$ are the number of neurons in the input and output layers, respectively.
Suppose ${\bf W}^l$ for $l=1,\ldots, L-1$ denotes the weight matrix of size $(n_l \times n_{l-1})$ associated with the $l$-th layer of the neural network, whose $(j,k)$-th entry, namely ${W}^l_{j,k}$, represents the weight connecting the $j$-th neuron in layer $l$ with the $k$-th neuron in layer $l-1$. 
Let ${\bf b}^l$ denote the bias vector in the $l$-th layer where $ {b}_j^l$ gives the bias associated with the $j$-th neuron in the $l$-th layer.
Let ${\bf Z}^l$ be the response observed from the $l$-th layer of the network before the activation. The response variable ${\bf Z}^l$ is then given by
\begin{align}\label{z_l}
{\bf Z}^l = {\bf W}^l ~ \text{NN}^{l-1}({\bf x},t,{\bf \phi}) + {\bf b}^l, ~~~ l=1,\ldots, L-1.
\end{align}
Suppose $\sigma$ is a nonlinear activation function applied element-wise to the response variable in the layers. 
The input layer is $\text{NN}^{0}({\bf x},t,{\bf \phi}) = {\bf x}$.
From \eqref{z_l}, the output from the $l$-th layer of the neural network is given by 
\begin{align}\label{nn_ll}
\text{NN}^{l}({\bf x},t,{\bf \phi}) = \sigma({\bf Z}^l), ~~~ l=1,\ldots, L-1
\end{align}
and the output in the last layer is as follows
\begin{align}\label{nn_L}
\text{NN}^{L}({\bf x},t,{\bf \phi}) = {\bf Z}^L\;.
\end{align}

Overall, the neural network architecture is characterized by its parameters (or weights) which can be encoded as follows ${\bf \phi}= ({\bf W}^1,\ldots,{\bf W}^L,{\bf b}^1,\ldots,{\bf b}^L)$.

To train a neural network, one requires the definition of a loss function, which measures the performance of the trained machine. Let $\mathcal{L}\left(\mathcal{N}^{L}({\bf x},t,{\bf \phi})\right)$ denote the loss function of the neural network $\mathcal{N}^{L}({\bf x},t,{\bf \phi})$.
In addition to the above forward propagation steps (see equations \eqref{nn_ll} and \eqref{nn_L}), one needs to consider the back-propagation to make sure that the responses in the network make sense. 
The back-propagation algorithm evaluates the gradients of the loss function with respect to the response variables in the previous layers. It is easy to show that the back-propagation boils down to 
\begin{eqnarray}
\label{back_pp_ll}
\frac{\partial \mathcal{L}\left(\mathcal{N}^{L}({\bf x},t,{\bf \phi})\right) }{\partial {\bf Z}^l} =
\left[
\left( {\bf W}^{l+1} \right)^\top \left( \frac{\partial \mathcal{L}\left(\mathcal{N}^{L}({\bf x},t,{\bf \phi})\right) }{\partial {\bf Z}^{l+1}} \right)
\right] \odot \frac{\partial \sigma({\bf Z}^l)}{\partial{\bf Z}^l}\; ,
\end{eqnarray}
where $\odot$ stands for the Hadamard product calculating the element-wise product between two matrices. Finally, one can compute the back-propagation of the last layer by
\begin{eqnarray}\label{back_pp_L}
\frac{\partial \mathcal{L}\left(\mathcal{N}^{L}({\bf x},t,{\bf \phi})\right) }{\partial {\bf Z}^L} =
\frac{\partial}{\partial \text{NN}^{L}({\bf x},t,{\bf \phi})} \mathcal{L}\left(\mathcal{N}^{L}({\bf x},t,{\bf \phi})\right) 
\odot \frac{\partial \sigma({\bf Z}^L)}{\partial{\bf Z}^L}.
\end{eqnarray}
For more details about the deep neural network algorithms, readers are referred to \cite{Goodfellow:Deep,Ramsundar}.

The Universal Approximation Theorem declares that neural networks are universal approximators: no matter the nature of a function, there is a neural network that can approximately approach its result. Because of this property of neural networks, various researchers have recently investigated ways to employ neural networks to solve ordinary and partial differential equations, e.g. \cite{Chen:2020}.

The basic concept in solving differential equations using an ANN is to re-frame the problem as an optimization problem in which the objective consists of minimizing the squared residual of the differential equations. To do so, one can generalize any differential equation as
\begin{equation*}
    \mathcal{D} u({\bf x},t) - f = 0\; ,
\end{equation*}
where $\mathcal{D}$ is the differential operator, $u({\bf x},t)$ is the solution of interest, and $f$ is a known forcing function. 

As described earlier, ANNs can be considered universal approximates for any function. We consider the output of a neural network as $\mathcal{N}^{L}({\bf x},t,{\bf \phi})$. $\mathcal{N}^{L}({\bf x},t,{\bf \phi})$ provides us with the approximation for our solution for the differential equation. Given $\mathcal{N}^{L}({\bf x},t,{\bf \phi})$ as a solution to the differential equation, we can construct the loss function by the squared residuals in the form of
\begin{equation}\label{loss_nn}
    \mathcal{L}(\mathcal{N}^{L}({\bf x},t,{\bf \phi})) =
    \left(\mathcal{N}^{L}({\bf x},t,{\bf \phi}) - f\right)^2 \; .
\end{equation}

The loss function \eqref{loss_nn} now can be incorporated into the forward and backward propagation steps \eqref{z_l}-\eqref{back_pp_L} to measure the loss corresponding to the set of learned parameters.

It is easily seen that equation \eqref{loss_nn} translates the problem of finding solutions to a system of differential equations, to an optimization process of the neural network loss function by
\begin{equation}\label{eq:nn_solver}
    \arg\min_{{\bf \phi}} \mathcal{L}(\mathcal{N}^{L}({\bf x},t,{\bf \phi})).
\end{equation}

Equation \eqref{eq:nn_solver} indicates that the neural network is designed to minimize the squared residuals of the trained $\mathcal{N}^{L}({\bf x},t,{\bf \phi})$. Consequently, this  guarantees that $\mathcal{N}^{L}({\bf x},t,{\bf \phi})$ represents the functional form of the solution of the differential equations when the neural network learning process converges.

We follow this approach in this work to tackle the problem of solving the system of ordinary differential equations defined by the equations of motion of black holes in the parabolic class, as we described in Section \ref{sec:equ_motion} (Equations \eqref{eq:b0} and \eqref{eq:f0''}). This system of equations can be approached with a neural network-based solver like the one described. Our goal here is to provide an alternative approach to General Relativity-based profile root-fining, described in Section \ref{sec:equ_motion}, so that we can explore whether there are self-similar solutions for high dimensions in the parabolic class. 

\section{Numerical Studies Based on ANNs}\label{sec:num_studies}
In this section, we explore the properties of the fully connected neural networks, described in Section \ref{sec:NNs}, in estimating the critical collapse functions corresponding to equations of motion in higher dimensional parabolic class. Using the neural network estimates for the critical collapse functions, we show that no self-similar critical solution exists for the parabolic class in higher dimensions of space-time.

Due to the invariance of $f(z)-{\bar f}(z)$, it is clear from the equations of motion that all the equations are invariant under a shift of $f(z)$. In other words, the entire equations do not depend on $u(z)$, implying that $u(0)=0$. On the other hand, the regularity of $\tau$ \eqref{eq:tauansatz} at $z=0$ implies $f'(0)=u'(0)+iv'(0)=0$. Hence, all boundary conditions are given by
\begin{equation}
   b(0)=1, ~ u'(0)=v'(0)=u(0)=0.
\label{881}
\end{equation}

Similar to Section \ref{sec:equ_motion}, from the regularity of $\tau$ and residual symmetries, the boundary conditions of the required critical functions are derived by \eqref{boundry_bf}.  In a similar vein, as the behaviours of the real and imaginary parts of $f$ do not depend on the value of $\Im f(0)$, we can set the boundary conditions of the two real parameters as $\Im f(0)=1$ and $\omega=1.65$ in the ANNs estimating method for five dimensional space-time. From \eqref{math12} we also set  $\omega =1.65$ and $\Im f(0)=0.321$ for all the other dimensions, see \cite{Hatefi:2020jdr,hatefialvarez1307}.

\begin{figure}
    \centering
    \includegraphics[width=4in]{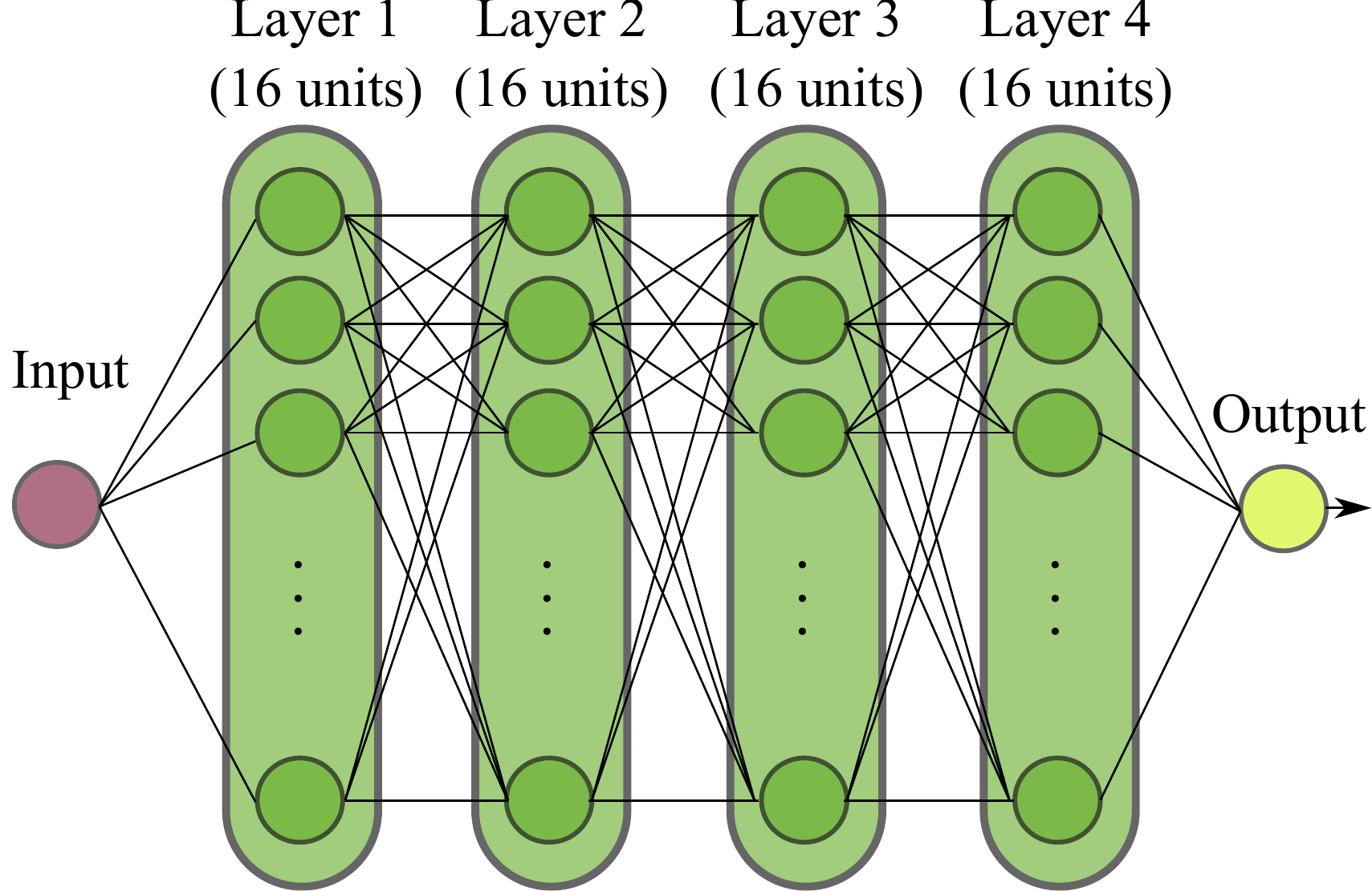}
    \caption{The ANN architecture implemented to optimize the parabolic equations of motion. It is a multi-layer perceptron of four hidden layers with 16 units each.}
    \label{fig:nn_model}
\end{figure}
 
To implement the ANN-based method, described in Section \ref{sec:NNs}, we used the python package NeuroDiffEq \cite{Chen:2020}. 
We have implemented a specific solver to obtain the estimates of the critical collapse functions $b_0(z)$, $Re(f(z))$ and $Im (f(z))$ corresponding to the parabolic class for $d=4,\ldots,9$ space-time dimensions. We integrate in this solver the NN network architecture shown in Figure \ref{fig:nn_model}. It consists of a multilayer perceptron for each function, with an input layer, a set of four hidden fully connected layers with 16 units each, and the output layer. Every hidden unit is followed by a rectified linear unit (ReLU) activation function \cite{Hahnloser, Goodfellow:Deep}. This architecture provides the necessary stability to find the solutions of the system of differential equations we are dealing with.

According to the profile root-finding studies discussed in Section \ref{sec:equ_motion}, if there was a parabolic solution to the equations of motion, the solution would appear in a neighbourhood of small values of space-time, namely $z < 3$. 
For this reason, using the above ANN configuration, we validated the estimates of the critical collapse functions at 1000 equally spaced points in the interval $z_j \in (0,3],~ j=1,\ldots,1000$. We then computed the ANN estimates using 3000 epochs to reach a good level of convergence of the loss functions in learning all the critical functions for $d=4,\ldots,9$ dimensions. 

Due to the non-linear patterns of the parabolic equations of motion, and the trained ANN solvers' sampling variability, the critical collapse function estimates slightly change from one trial to another. To cope with this stochastic property of the proposed ANN estimates, we independently replicated 20 times the above ANN estimation procedure. That is, the ANN solver was independently applied to the differential equations for 20 times. We then validated the performance of the trained ANN models at points $z_j, ~ j=1,\ldots,1000$. Skewness in the performance of the ANN estimations is unavoidable because of the non-linear pattern of the critical functions in their domain and the stochastic property of the learned parameters.
Therefore, the critical collapse functions in the domain are considered our statistical parameters of interest. We then constructed the 95\% empirical confidence intervals for all the critical collapse functions. To construct the confidence intervals, we obtained the 2.5, 50 and 97.5 percentiles of the ANN-based estimates at each space-time point $z_j, j=1,\ldots,1000$. We finally replicated the above ANN-based method to estimate the parabolic critical collapse functions in all dimensions $d=4,\ldots,9$. 

\begin{figure}
    \centering
    \includegraphics[width=4.5in]{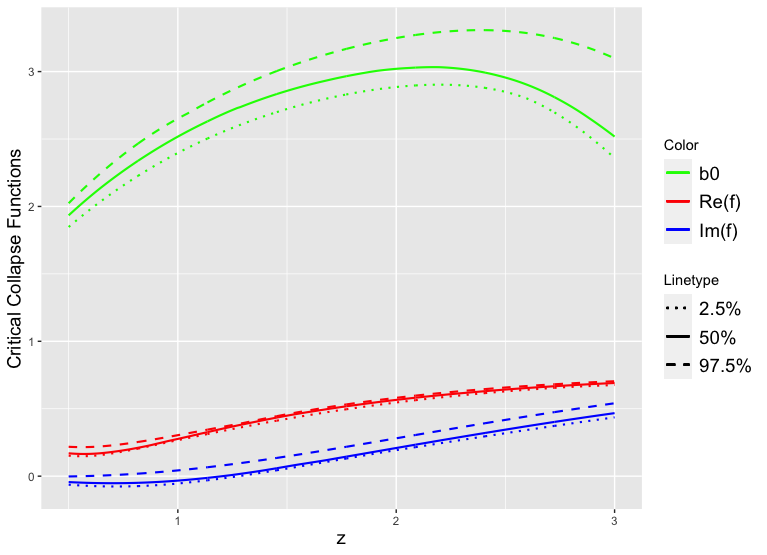}
    \caption{The 95\% ANN-based confidence intervals for the critical collapse functions in the four-dimensional parabolic class. The ANN lower (dotted lines), upper (dashed lines) and median (solid lines) bands  were computed from 20 replicates with 3000 epochs each. }
    \label{NN_4d}
\end{figure}

\begin{figure}
    \centering
    \includegraphics[width=4.5in]{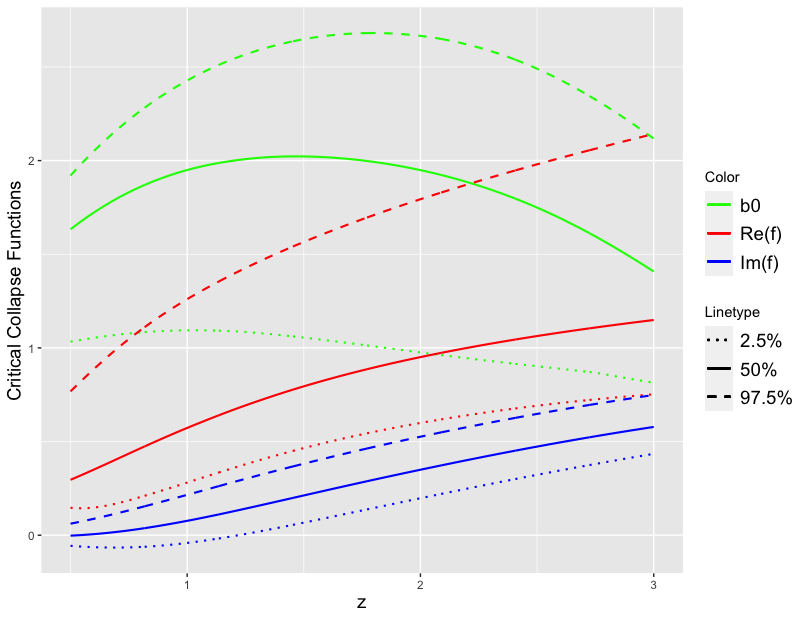}
    \caption{The 95\% ANN-based confidence intervals for the critical collapse functions in the five-dimensional parabolic class. The ANN lower (dotted lines), upper (dashed lines) and median (solid lines) bands were computed from 20 replicates with 3000 epochs each.}
    \label{NN_5d_20sim}
\end{figure} 

\begin{figure}
    \centering
    \includegraphics[width=6in]{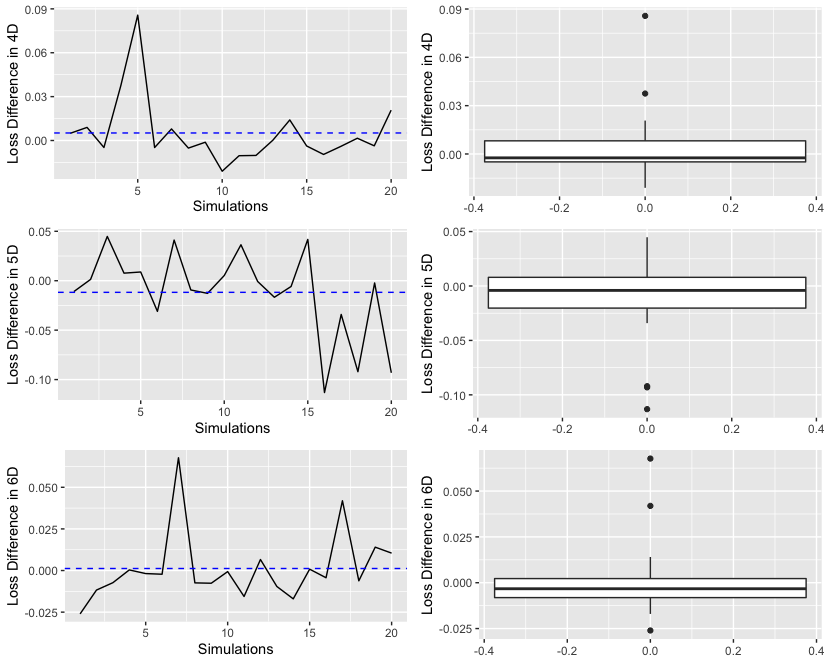}
    \caption{The differences between train and validation loss functions of the developed ANNs in estimating the critical collapse functions. The top, middle and bottom panels show the loss differences of the estimates in 4D, 5D and 6D parabolic spaces, respectively. The panels show the line plot and box plot of loss differences of the ANN estimates for 20 replicates with 3000 epochs each.}
    \label{R_d456box}
\end{figure}

\begin{figure}
    \centering
    \includegraphics[width=4.5in]{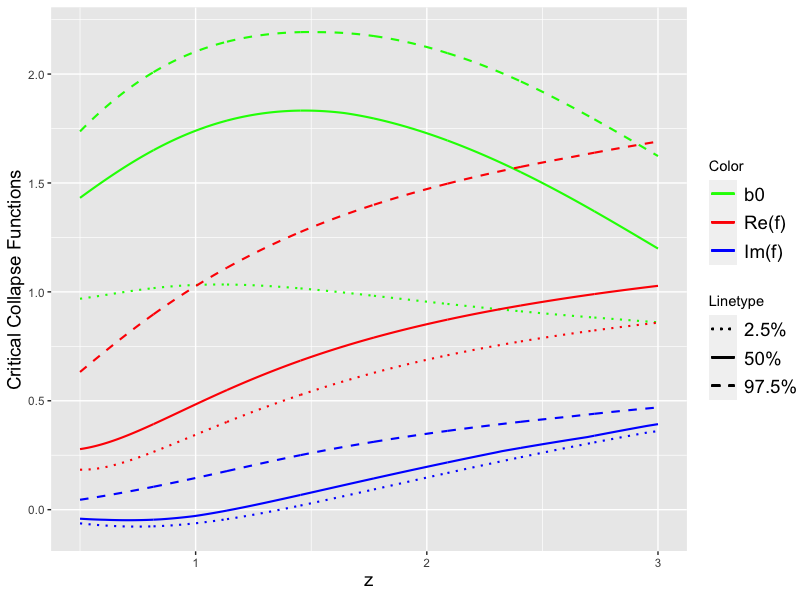}
    \caption{The 95\% ANN-based confidence intervals for the critical collapse functions in the six-dimensional parabolic class. The ANN lower (dotted lines), upper (dashed lines) and median (solid lines) bands  were computed from 20 replicates with 3000 epochs each.}
    \label{NN_6d}
\end{figure}

\begin{figure}[H]
    \centering
    \includegraphics[width=4.5in]{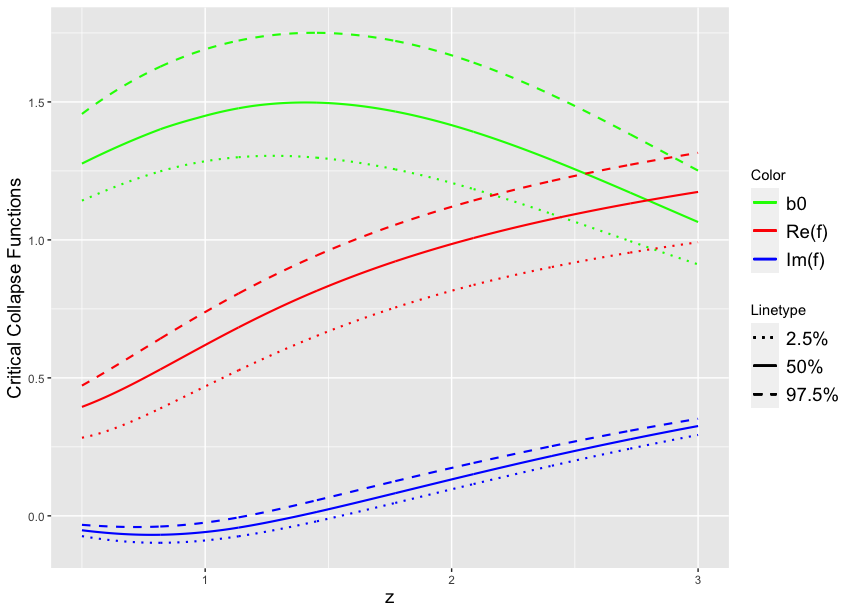}
    \caption{The 95\% ANN-based confidence intervals for the critical collapse functions in the seven-dimensional parabolic class. The ANN lower (dotted lines), upper (dashed lines) and median (solid lines) bands  were computed from 20 replicates with 3000 epochs each.}
    \label{NN_7d}
\end{figure}

\begin{figure}[H]
    \centering
    \includegraphics[width=4.5in]{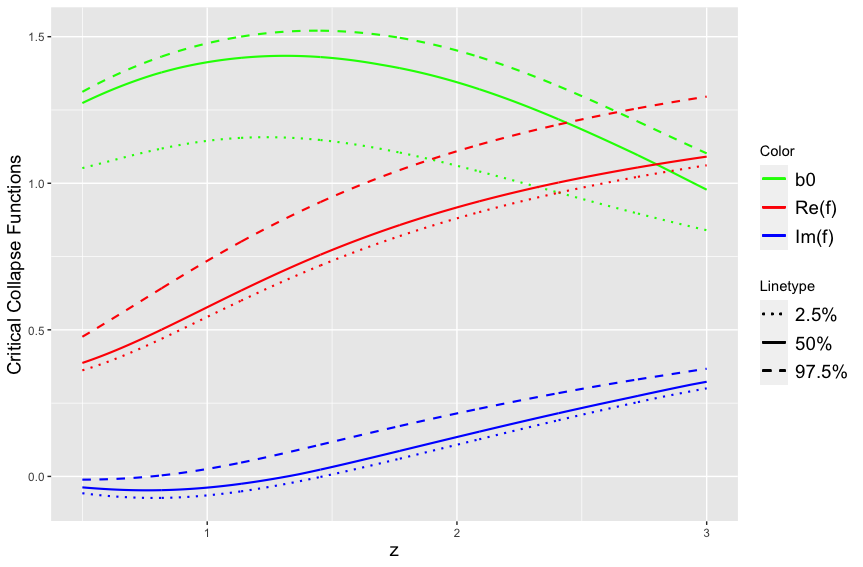}
    \caption{The 95\% ANN-based confidence intervals for the critical collapse functions in the eight-dimensional parabolic class. The ANN lower (dotted lines), upper (dashed lines) and median (solid lines) bands  were computed from 20 replicates with 3000 epochs each.}
    \label{NN_8d}
\end{figure}

\begin{figure}[H]
    \centering
    \includegraphics[width=4.5in]{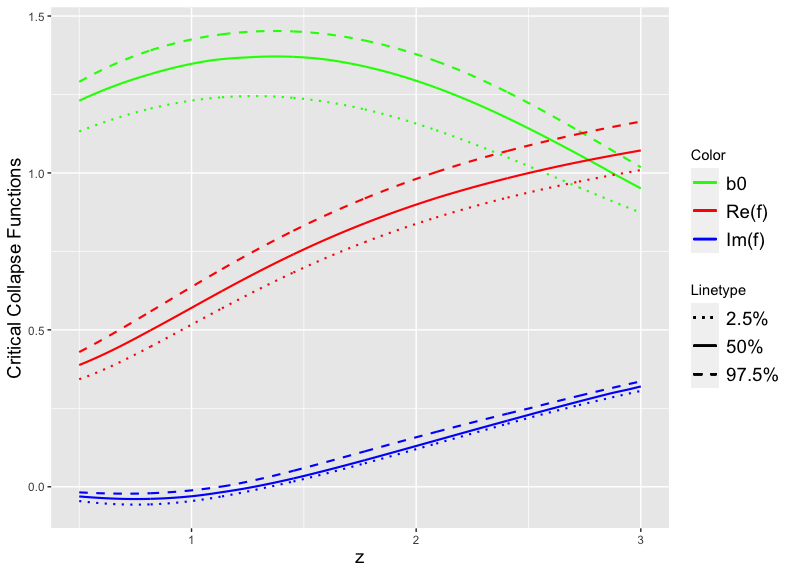}
    \caption{The 95\% ANN-based confidence intervals for the critical collapse functions in the nine-dimensional parabolic class. The ANN lower (dotted lines), upper (dashed lines) and median (solid lines) bands  were computed from 20 replicates with 3000 epochs each.}
    \label{NN_9d}
\end{figure}

Figures \ref{NN_4d} - \ref{NN_9d} show the 95\% ANN-based confidence intervals of the critical collapse functions $b_0(z)$ (green colour), $Re(f(z))$ (red colour) and $Im (f(z))$ (blue colour) corresponding to parabolic class for $d=4,\ldots,9$ space-time dimensions. The lower, upper and middle bands of the confidence intervals are represented by 2.5  (dotted line), 97.5 (dashed line) percentiles and median (solid line), respectively.  
We observe from Figures \ref{NN_4d} - \ref{NN_9d} that no intersection occurs between 95\% ANN-based confidence regions of functions $Re(f(z))$ and $Im (f(z))$ in parabolic class of $d=4,\ldots,9$ dimensions.  
As the dimension grows, in addition, more divergence is observed between the ANN-based confidence regions of the real and imaginary parts of $G(\omega,\Im f(0))$. 
Therefore, Figures \ref{NN_4d} - \ref{NN_9d} confirm that, at 95\% level of confidence, there exists no self-similar black hole solution for parabolic class in $d=4,\ldots,9$ dimensions. This ANN-based estimation method confirms the numerical results of the profile root-finding of Section \ref{sec:equ_motion}, but from a different perspective, directly tackling the system of differential equations defined by the equations of motion of black holes.

\begin{figure}[H]
    \centering
    \includegraphics[width=6in]{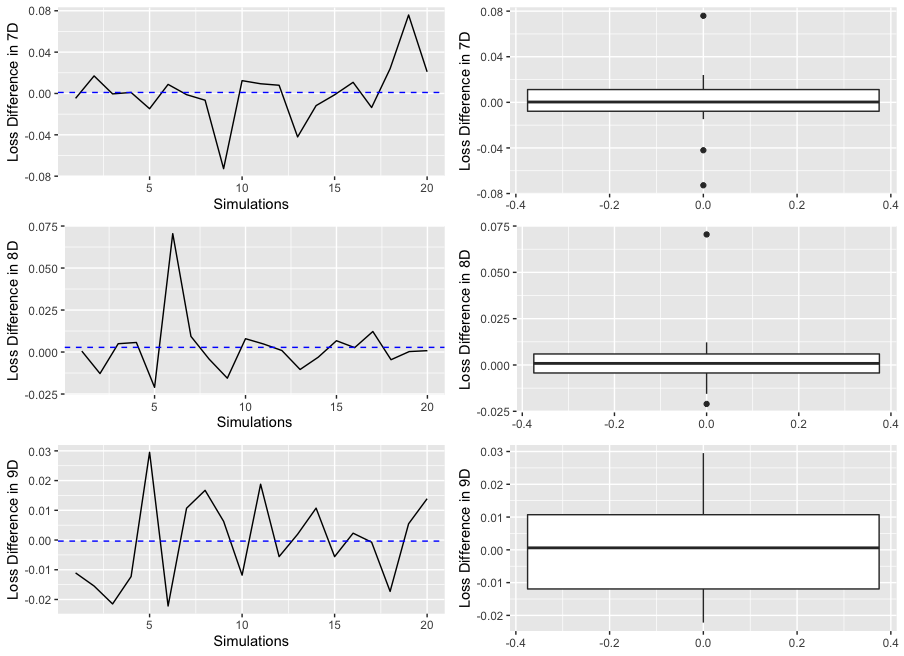}
    \caption{The differences between train and validation loss functions of the developed ANNs in estimating the critical collapse functions. The top, middle and bottom panels show the loss differences  of the estimates in 7d, 8d and 9d parabolic spaces, respectively. The panels show the line plot and box plot of loss differences of the ANN estimates for 20 replicates with 3000 epochs each.}
    \label{R_d789box}
\end{figure}

We assess the convergence of the ANN estimators for the critical collapse functions in Figures \ref{R_d456box} and \ref{R_d789box}. With respect to the effect of the learned parameters (for the ANN) on the behaviour of the loss functions, we analyzed the loss differences observed in the 20 replicates of the ANN estimation method. To do so, we computed the difference between training and test loss functions in the last epoch of the ANN learning process. Figures \ref{R_d456box} and \ref{R_d789box} show the line and box plots of this loss difference of the proposed ANN method in the estimation of the critical collapse functions. 
We clearly observe that the median difference between training and test losses is close to zero. Despite the sampling variability, this confirms, on average, the convergence of the ANN-based method in solving the equations of motion in $d=4,\ldots,9$ dimensions for parabolic class.

In Figure \ref{NN_5d_20sim} there is a short intersection between the 95\% ANN-based confidence intervals of the real and imaginary parts of $G(\omega,\Im f(0))$ in five dimensions. This spurious black hole solution is compatible with the finding of Hatefi and Vanzan \cite{Hatefi:2020jdr} where equations of motion in the five-dimensional parabolic class may sometimes lead to a solution due to the numerical instability of the solvers. 

\begin{figure}[H]
    \centering
    \includegraphics[width=4.5in]{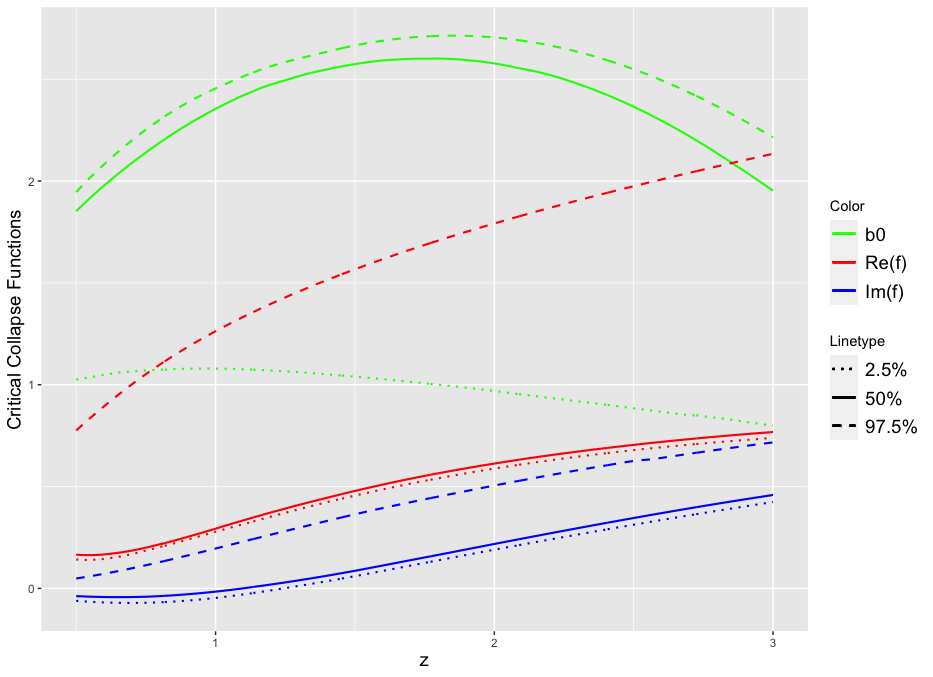}
    \caption{The 95\% ANN-based confidence intervals for the critical collapse functions in the five-dimensional parabolic class. The ANN lower (dotted lines), upper (dashed lines) and median (solid lines) bands  were computed from 100 replicates with 3000 epochs each.}
    \label{NN_5d_sim100_ep3000}
\end{figure}

\begin{figure}[H]
    \centering
    \includegraphics[width=6in]{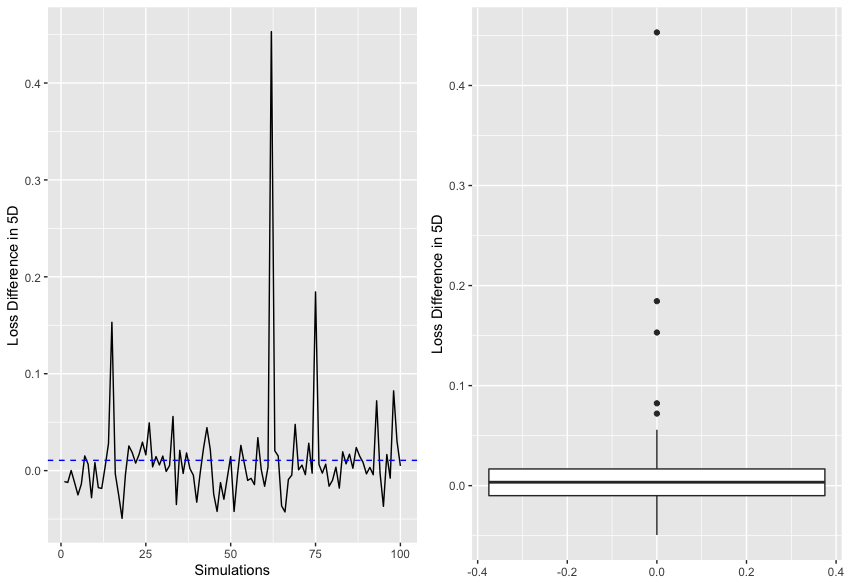}
    \caption{The differences between train and validation loss functions of the developed ANNs in estimating the critical collapse functions in 5d parabolic class. The panels show the line plot and box plot of loss differences of the ANN estimates for 100 replicates with 3000 epochs each.}
    \label{NN_box_5d_sim100}
\end{figure}

To elaborate on this argument, we investigated another numerical study where more estimates are considered. Here, we followed our ANN method, as described earlier, in estimating the critical functions in five dimensions with 3000 epochs. We replicated the ANN estimates 100 times instead of just 20 times. Figures \ref{NN_5d_sim100_ep3000} and \ref{NN_box_5d_sim100} show the results of this extensive numerical study. From Figure \ref{NN_5d_sim100_ep3000}, we observe that there is no intersection between 95\% ANN-confidence intervals of the real and imaginary parts of  $G(\omega,\Im f(0))$ in 5 dimensions. Therefore, at a 95\% confidence level, no parabolic black hole solution exists even in 5 dimensions. 
We believe that the spurious solution is due to the complexity of the equations of motion and the numerical instability of the neural network solvers. The ANN solver sometimes converges to the local optimum rather than the global optimum. Therefore the variance of the ANN statistic increases. This numerical instability consequently increases the length of the ANN-confidence intervals of the critical collapse functions. Figure \ref{NN_box_5d_sim100} shows clearly this numerical instability such that the ANN solver converges to a local optimum of the equations of motion for 3-4 replicates (out of 100 replicates) and results in a significant loss difference. 

\section{Summary and Concluding Remarks}\label{sec:sum}

In this paper, we proposed a numerical approach to find the self-similar solutions to the axion-dilaton system in higher dimensions of the parabolic class. Specifically, we have described an ANN-based approach that allows us to directly solve the differential equations proposed in the axion-dilaton model. Unlike the statistical models in \cite{Hatefi:2022shc, Hatefi:2021xwh}, the ANN-based estimators take directly the equations of motion as input and translate the estimation problem into an optimization process. The proposed numerical method is quite generic and can be applied to all SL(2, R) transformation classes in any dimension and any matter content. 
In fact, we have shown in Appendix \ref{sec:appendix} that the proposed ANN-based approach could confirm that there exists a solution which is compatible with the findings in  the literature for the four-dimensional elliptic case. 
We found that the proposed ANNs estimation approach detected a solution in the equation of motions with probability 1 in the numerical experiment. This empirically confirms a solution in the domain of the critical collapse functions in the elliptic class of four dimensions.
  
Unlike elliptic and hyperbolic classes, for the parabolic class in higher dimensional space-time we have shown that there exists no self-similar critical solution due to an extra scaling symmetry. 
According to the sampling variability of neural networks, we constructed  95\% empirical confidence intervals based on our fully connected ANNs for all the critical collapse functions in parabolic class for various dimensions ranging from four to nine. Through extensive numerical studies, at a 95\% confidence level, the ANN-based solvers confirm that there is no black hole solution in higher dimensions for the parabolic class.

The gravitational collapse does not thus take place for parabolic ansatz. Our results provide some doubts about the universality of the Choptuik phenomena. Therefore, the fastest-growing mode of the perturbations determining the Choptuik exponent does not exist in these cases, and no gravitational collapse occurs. 
It is also worth mentioning that our analysis provides reasons not to expect to transfer the standard derivations of statistical mechanics to the critical gravitational collapse.
 

\section*{Acknowledgments}

The authors would like to thank the editor and the anonymous referee for their constructive comments which improved the quality and presentation of the manuscript. Ehsan Hatefi would like to thank  A. Kuntz,  E. Hirschmann, C. Gutiérrez-Álvarez and L. Álvarez-Gaumé for fruitful conversations. The research of Ehsan Hatefi is supported by the María Zambrano Grant of the Ministry of Universities of Spain. Armin Hatefi acknowledges the support from the Natural Sciences and Engineering Research Council of Canada (NSERC).

\appendix
\newpage 
\section{Solution for the Elliptic Class in 4 dimension}
\label{sec:appendix}
Although we planned to develop the ANNs throughout this manuscript to show no black hole solution exists in higher dimensional space-time of the parabolic class, it is instructive to assess if the proposed ANN-based method can detect if there exists a solution to equations of motion in the case where the solution is already known in the literature. To do so, we focus on the elliptic space-time in 4 dimension. In the following, we briefly describe the equations of motion corresponding to the elliptic class of 4 dimension.

In a similar vein to Section \ref{sec:equ_motion}, applying CSS ans\"atze \eqref{eq:tauansatz} to equations of motion \eqref{eq:efes} and \eqref{eq:taueom}, one can derive the ordinary differential equations for $u(z)$, $b(z)$, $f(z)$. Due to the spherical symmetry, $u(z)$ and $u'(z)$ can be eliminated. From \eqref{eq:efes} and \eqref{eq:taueom}, the equations of motion for self-similar solutions can be re-written as:
\begin{eqnarray}
0 & = & b' + { z(b^2 - z^2) \over b (-1 + |f|^2)^2} f' \bar{f}' - {
i \omega (b^2 - z^2) \over b (-1 + |f|^2)^2} (f \bar{f}' - \bar{f} f')
- {\omega^2 z |f|^2 \over b (-1 + |f|^2)^2}, \nonumber\\
0 & = & f''
     - {z (b^2 + z^2) \over b^2 (-1 + |f|^2)^2} f'^2 \bar{f}'
     + {2 \over (1 - |f|^2)} \left(1
       - {i \omega (b^2 + z^2) \over 2b^2 (1 - |f|^2)} \right) \bar{f} f'^2 \nonumber \\&&
     + {i \omega (b^2 + 2 z^2) \over b^2 (-1 + |f|^2)^2} f f'
\bar{f}' 
  + {2 \over z} \left(1 + {i \omega z^2 (1 + |f|^2) \over (b^2 - z^2)
(1 - |f|^2)}\right.\nonumber \\&& 
+ \left.{\omega^2 z^4 |f|^2 \over b^2 (b^2 - z^2) (1 -
|f|^2)^2}\right) f'+ {\omega^2 z \over b^2 (-1 +|f|^2)^2} f^2
\bar{f}' + \nonumber \\&&
{2i \omega \over (b^2 - z^2)} \left(\frac{1}{2} - {i \omega (1 + |f|^2)
\over 2(1 - |f|^2)}\right.
- \left.{\omega^2 z^2 |f|^2 \over 2b^2 (-1 + |f|^2)^2}
\right) f.
\label{1fzeom321}
\end{eqnarray}

The above equations are invariant under a global redefinition of the phase of $f(z)$.  One can solve them in polar coordinates, that is $f(z)=f_m(z) e^{if_a(z)}$. The vanishing of the divergent part of $f''(z)$ generates a complex-valued constraint at $z_+$, denoted by  $G(b(z_+), f(z_+), f'(z_+)) = 0$. The explicit form of the $G$ function for the elliptic case in 4d is then given by:
\begin{eqaed}\label{eq:Gelliptic}
     G(f(z_+),f'(z_+)) = & \, 2 z \bar{f}(z_+) \left(-2 \omega^2\right) f'(z_+)\\ & +f(z_+) \bar{f}(z_+) \left(2 z_+ \bar{f}(z_+) (-2+2 i \omega+2) f'(z_+)+2 i \omega \left(2+\omega^2\right)\right)\\&-\frac{2 z_+ (2+2 i \omega-2) f'(z_+)}{f(z_+)}+2\omega (\omega-i) f(z_+)^2 \bar{f}(z_+)^2-2 \omega (\omega+i)\,.
\end{eqaed}

The regularity of $\tau$ and residual symmetries in the equations of motions \eqref{1fzeom321} result in the initial boundary conditions as $b(0) = 1,   f_m'(0) =f_a'(0)=f_a(0)=0$.

The behaviours of equations in \eqref{1fzeom321} do not depend on the value of phase of $f(z)$; hence, one can easily set the boundary conditions as $f_a(0) = 0$. Similar to Section \ref{sec:equ_motion}, applying the profile root-finding, one can easily obtain $\omega = 1.176$  and $f_m(0) = 0.892$ for four-dimensional space-time. Like in Section \ref{sec:equ_motion}, if there is an elliptic solution to the equations of motion, the solution should appear in a neighbourhood of small space-time values, namely $z < 3$. In this Appendix, we present a numerical study to show if the proposed ANNs approach can detect the solution in the equations of motion for the elliptic case in 4d.

Similar to numerical studies of Section \ref{sec:num_studies},  we validated the estimates of the critical collapse functions at 1000 equally spaced points in the interval $z_j = (0, 4]$ , $j = 1, \ldots, 1000$, using the same ANN architecture. We then computed the ANN estimates using 1000 epochs to reach a good level of convergence for the loss functions in learning all the critical functions for $d = 4$ dimension.

As discussed in Section \ref{sec:num_studies}, the ANNs require a training step to propagate and back-propagate the gradients for training the parameters; hence the ANN-based critical collapse functions change slightly from one trial to another, depending on the trained parameters. To account for this stochastic behaviour, we ran the ANN method with 1000 epochs in estimating the critical collapse functions, and then we replicated ANN estimation for $N = 20$ times (as in our previous study for the parabollic class). 

\begin{figure}
    \centering
    \includegraphics[width=4.5in]{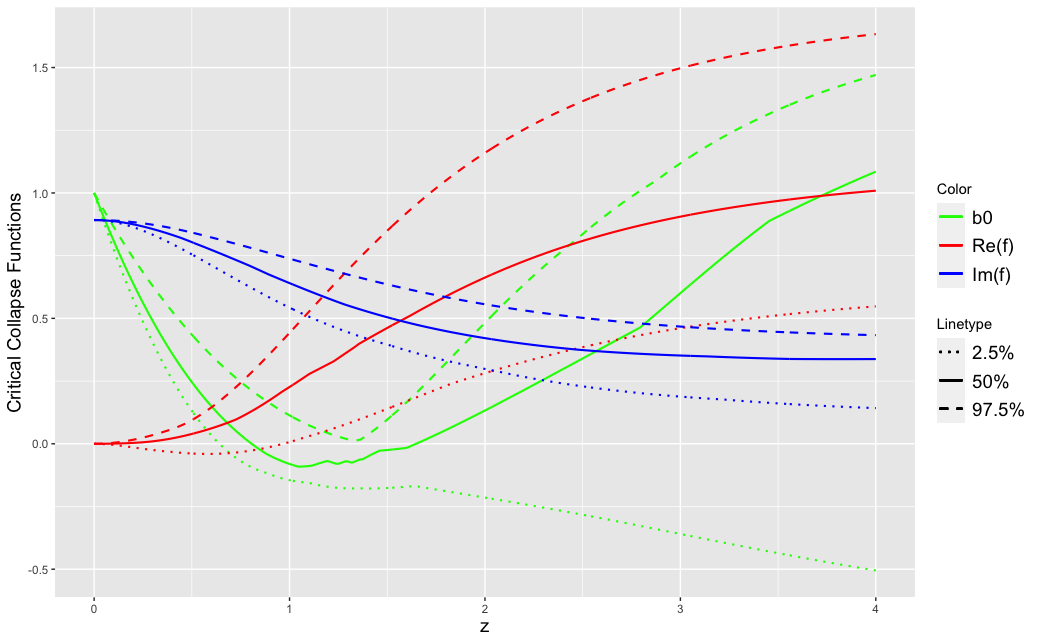}
        \caption{The 95\% ANN-based confidence intervals for the critical collapse functions in the four-dimensional elliptic class. The ANN lower (dotted lines), upper (dashed lines) and median (solid lines) bands  were computed from 20 replicates with 1000 epochs each.}    \label{elp_sim20_CI}
\end{figure}

\begin{figure}
    \centering
    \includegraphics[width=4.5in]{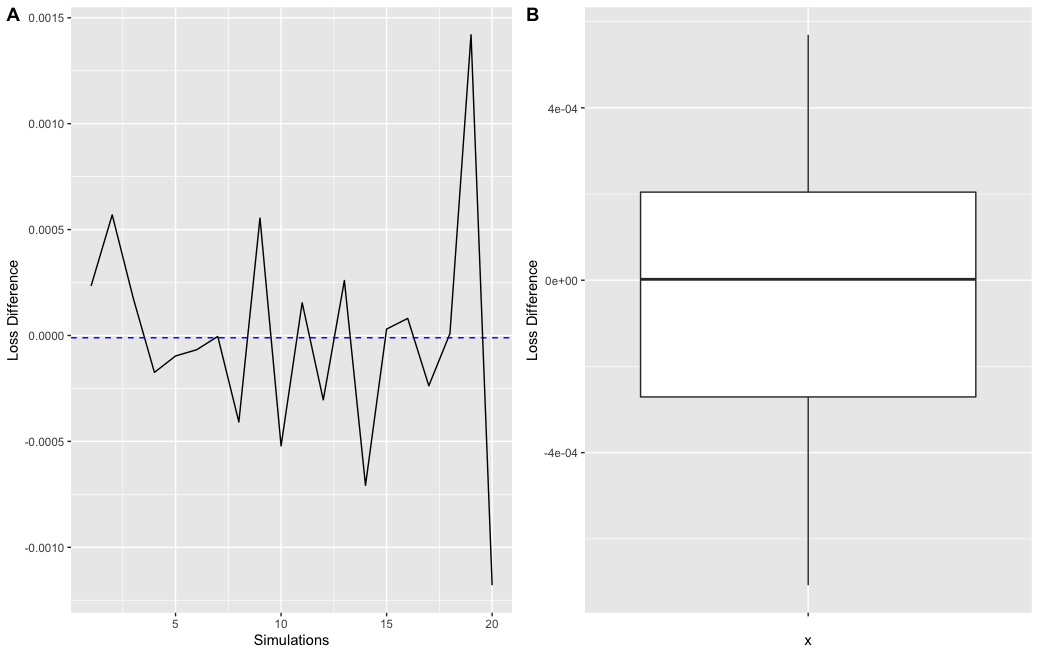}
    \caption{The differences between train and validation loss functions of the developed ANNs in estimating the critical collapse functions in 4d elliptic class. The panels show the line plot and box plot of loss differences of the ANN estimates for 20 replicates.}
    \label{elp_sim20_box}
\end{figure}

We used three statistical measures to evaluate the performance of the ANNs in estimating the critical collapse functions. Like in Section \ref{sec:num_studies}, we first developed the 95\%  ANN-based confidence interval for the critical collapse functions in elliptic space. Figure \ref{elp_sim20_CI} represents the ANN-based confidence intervals based on 20 replicates. We also computed the difference between training and validation losses in the last epoch of the ANNs. Figure \ref{elp_sim20_box} represents the line and box plots of the loss differences over 20 replicates. Finally, we computed the coverage probability to empirically measure the likelihood that the ANNs approach captures a solution in the domain of the critical collapse functions. To compute the coverage probability, when ANNs detect a solution in replicate $i$, we set the coverage index $I_i=1$; otherwise, $I_i=0$ for $i=1,\ldots,N$. Finally, the coverage probability is calculated by $\frac{1}{N}\sum_{i=1}^{N} I_i$. 

Figure \ref{elp_sim20_box} shows that the difference between training and validation loss functions is almost $10^{-3}$. This indicates that with 1000 epochs the ANNs have converged in estimating the critical collapse functions in the elliptic space-time of 4d. From Figure \ref{elp_sim20_CI}, it is seen that the ANNs confirm, at the 95\% confidence level, that there is a solution in the elliptic class of 4d roughly between $z_+ \in [0.5,3]$, which is compatible with the analytical solution $z_+=2.605$ from \cite{Antonelli:2019dqv}. Finally, we 
observed that the proposed ANNs estimation approach detected a solution in the equation of motions with coverage probability 1 in the numerical experiment. This empirically confirms the solution in the domain of the critical collapse functions in the elliptic class of 4d.

\begin{thebibliography}{2007}
\bibitem{Chop} M.W.~Choptuik,
{ Universality and Scaling in Gravitational Collapse of a Massless Scalar Field,}
{\it Phys.\ Rev. Lett.}\ {\bf70}, 9 (1993).
\bibitem{Christodolou}
   D.~Christodoulou,
  ``The Problem of a Self-gravitating Scalar Field,''
  Commun.\ Math.\ Phys.\  {\bf 105} (1986) 337; ``Global Existence of Generalized Solutions of the Spherically Symmetric Einstein Scalar Equations in the Large,''
  Commun.\ Math.\ Phys.\  {\bf 106} (1986) 587;``The Structure and Uniqueness of Generalized Solutions of the Spherically Symmetric Einstein Scalar Equations,''
  Commun.\ Math.\ Phys.\  {\bf 109} (1987) 591.
\bibitem{Hamade:1995ce}
  R.~S.~Hamade and J.~M.~Stewart,
  ``The Spherically symmetric collapse of a massless scalar field,''
  Class.\ Quant.\ Grav.\  {\bf 13} (1996) 497
  [arXiv:gr-qc/9506044].
\bibitem{KHA} T.~Koike, T.~Hara, and S.~Adachi,
``Critical Behavior in Gravitational Collapse of Radiation Fluid:
A Renormalization Group (Linear Perturbation) Analysis,''
Phys.\ Rev.\ Lett.\  {\bf 74} (1995) 5170
  [gr-qc/9503007].
\bibitem{AlvarezGaume:2006dw}
  L.~Alvarez-Gaume, C.~Gomez and M.~A.~Vazquez-Mozo,
  ``Scaling Phenomena in Gravity from QCD,''
  Phys.\ Lett.\ B {\bf 649} (2007) 478
  [hep-th/0611312].
\bibitem{Birukou:2002kk}
  M.~Birukou, V.~Husain, G.~Kunstatter, E.~Vaz and M.~Olivier,
  ``Scalar field collapse in any dimension,''
  Phys.\ Rev.\ D {\bf 65} (2002) 104036
  [gr-qc/0201026].

\bibitem{Sorkin:2005vz}
  E.~Sorkin and Y.~Oren,
  ``On Choptuik's scaling in higher dimensions,''
  Phys.\ Rev.\  D {\bf 71}, 124005 (2005)
  [arXiv:hep-th/0502034].


  
\bibitem{Bland:2005vu}
  J.~Bland, B.~Preston, M.~Becker, G.~Kunstatter and V.~Husain,
  ``Dimension-dependence of the critical exponent in spherically symmetric gravitational collapse,''
  Class.\ Quant.\ Grav.\  {\bf 22} (2005) 5355
  [gr-qc/0507088].

\bibitem{HirschmannEardley}
  E.~W.~Hirschmann and D.~M.~Eardley,
  ``Universal scaling and echoing in gravitational collapse of a complex scalar field,'' Phys.\ Rev.\ D {\bf 51} (1995) 4198
  [gr-qc/9412066].
\bibitem{Rocha:2018lmv}
J.~V.~Rocha and M.~Toma\v{s}evi\'c,
``Self-similarity in Einstein-Maxwell-dilaton theories and critical collapse,''
Phys. Rev. D \textbf{98} (2018) no.10, 104063
[arXiv:1810.04907 [gr-qc]].
\bibitem{AlvarezGaume:2008qs}
  L.~Alvarez-Gaume, C.~Gomez, A.~Sabio Vera, A.~Tavanfar and M.~A.~Vazquez-Mozo,``Critical gravitational collapse: towards a holographic understanding of the Regge region,''
  Nucl.\ Phys.\ B {\bf 806} (2009) 327
  [arXiv:0804.1464 [hep-th]].
\bibitem{evanscoleman}
  C.~R.~Evans and J.~S.~Coleman,
  ``Observation of critical phenomena and selfsimilarity in the gravitational collapse of radiation fluid,''
  Phys.\ Rev.\ Lett.\  {\bf 72} (1994) 1782
  [gr-qc/9402041].
  \bibitem{MA} D.~Maison,
``Non-Universality of Critical Behaviour in Spherically Symmetric
Gravitational Collapse,''
Phys.\ Lett.\ B {\bf 366} (1996) 82
  [gr-qc/9504008].
\bibitem{Strominger:1993tt}
  A.~Strominger and L.~Thorlacius,
  ``Universality and scaling at the onset of quantum black hole formation,''
  Phys.\ Rev.\ Lett.\  {\bf 72} (1994) 1584
  [hep-th/9312017].
\bibitem{Hatefi:2020gis}
E.~Hatefi and A.~Kuntz,
``On Perturbation Theory and Critical Exponents for Self-Similar Systems,''
Eur. Phys. J. C \textbf{81} (2021) no.1, 15
[arXiv:2010.11603 [hep-th]].


  
\bibitem{Hirschmann:1995qx}
  E.~W.~Hirschmann and D.~M.~Eardley,
  `Critical exponents and stability at the black hole threshold for a complex scalar field,''
  Phys.\ Rev.\ D {\bf 52} (1995) 5850
  [gr-qc/9506078].
\bibitem{AE}
  A.~M.~Abrahams and C.~R.~Evans,
  ``Critical behavior and scaling in vacuum axisymmetric gravitational collapse,''
  Phys.\ Rev.\ Lett.\  {\bf 70} (1993) 2980.
\bibitem{AlvarezGaume:2008fx}
  L.~Alvarez-Gaume, C.~Gomez, A.~Sabio Vera, A.~Tavanfar and M.~A.~Vazquez-Mozo,
  ``Critical formation of trapped surfaces in the collision of gravitational
  shock waves,''
  JHEP {\bf 0902}, 009 (2009)
  [arXiv:0811.3969 [hep-th]].
\bibitem{Hirschmann_1997}
  E.~W.~Hirschmann and D.~M.~Eardley,
  ``Criticality and bifurcation in the gravitational collapse of a selfcoupled scalar field,''
  Phys.\ Rev.\ D {\bf 56} (1997) 4696
  [gr-qc/9511052].




\bibitem{Maldacena:1997re}
J.~M.~Maldacena,
``The Large N limit of superconformal field theories and supergravity,''Int. J. Theor. Phys. \textbf{38} (1999), 1113-1133, Adv. Theor. Math. Phys. \textbf{2}, arXiv:hep-th/9711200, E.~Witten, 
``Anti-de Sitter space and holography,''
Adv. Theor. Math. Phys. \textbf{2} (1998), 253-291,hep-th/9802150, S.~Gubser, I.~R.~Klebanov and A.~M.~Polyakov,
``Gauge theory correlators from noncritical string theory,''
Phys. Lett. B \textbf{428} (1998), 105-114, hep-th/9802109.

\bibitem{Birmingham:2001hc}
D.~Birmingham,
``Choptuik scaling and quasinormal modes in the AdS / CFT correspondence,''
Phys. Rev. D \textbf{64} (2001), 064024
[arXiv:hep-th/0101194 [hep-th]].



\bibitem{Hatefi:2012bp}
E.~Hatefi, A.~Nurmagambetov and I.~Park,
``ADM reduction of IIB on $\mathcal{H}^{p,q}$ to dS braneworld,''
JHEP \textbf{04} (2013), 170, arXiv:1210.3825 ,
``$N^3$ entropy of $M5$ branes from dielectric effect,''
Nucl. Phys. B \textbf{866} (2013), 58-71, arXiv:1204.2711,
S.~de Alwis, R.~Gupta, E.~Hatefi and F.~Quevedo,
``Stability, Tunneling and Flux Changing de Sitter Transitions in the Large Volume String Scenario,''
JHEP \textbf{11} (2013), 179, arXiv:1308.1222,

\bibitem{Ghodsi_2010}
  A.~Ghodsi and E.~Hatefi,
  ``Extremal rotating solutions in Horava Gravity,''
  Phys.\ Rev.\ D {\bf 81} (2010) 044016
  [arXiv:0906.1237 [hep-th]].


\bibitem{Hamade:1995jx}
  R.~S.~Hamade, J.~H.~Horne and J.~M.~Stewart,
  ``Continuous Self-Similarity and $S$-Duality,''
  Class.\ Quant.\ Grav.\  {\bf 13} (1996) 2241
  [arXiv:gr-qc/9511024].

\bibitem{AlvarezGaume:2011rk}
  L.~Álvarez-Gaumé and E.~Hatefi,
  ``Critical Collapse in the Axion-Dilaton System in Diverse Dimensions,'' Class.\ Quant.\ Grav.\  {\bf 29} (2012) 025006
  [arXiv:1108.0078 [gr-qc]].

\bibitem{hatefialvarez1307}
  L.~Álvarez-Gaumé and E.~Hatefi,
  ``More On Critical Collapse of Axion-Dilaton System in Dimension Four,''
  JCAP {\bf 1310} (2013) 037
  [arXiv:1307.1378 [gr-qc]].


\bibitem{ours}
  R.~Antonelli and E.~Hatefi,
  ``On self-similar axion-dilaton configurations,''
, JHEP \textbf{03} (2020), 074
[arXiv:1912.00078 [hep-th]].




\bibitem{Antonelli:2019dqv}
R.~Antonelli and E.~Hatefi,
``On Critical Exponents for Self-Similar Collapse,''
JHEP \textbf{03} (2020), 180
[arXiv:1912.06103 [hep-th]].


\bibitem{Hatefi:2022shc}
E.~Hatefi and A.~Hatefi,
``Nonlinear statistical spline smoothers for critical spherical black hole solutions in 4-dimension,''
Annals Phys. \textbf{446} (2022), 169112
[arXiv:2201.00949 [gr-qc]].

\bibitem{Hatefi:2021xwh}
E.~Hatefi and A.~Hatefi,
``Estimation of Critical Collapse Solutions to Black Holes with Nonlinear Statistical Models,''
Mathematics \textbf{10} (2022) no.23, 4537
[arXiv:2110.07153 [gr-qc]].





 

\bibitem{Chen:2020}
F. Chen, D. Sondak, P. Protopapas, P.Mattheakis, M. Liu, S.Agarwal, D. \& Di Giovanni, M.,
``NeuroDiffEq: A Python package for solving differential equations with neural networks,''  Journal Of Open Source Software. \textbf{5}, 1931 (2020)




\bibitem{DeepXDE}
  Lu. Lu, Xuhui. Meng, Zhiping. Mao, and George Em. Karniadakis,
``DeepXDE: A Deep Learning Library for Solving Differential Equations,'' SIAM Review, \textbf{63}, 208-228 (2021)

 \bibitem{pydens_2019}
  Alexander Koryagin, Roman Khudorozkov, and Sergey Tsimfer
``PyDEns framework for solving differential equations with deep learning,'' 2019



 


  
\bibitem{Sen:1994fa}
  A.~Sen,
  ``Strong - weak coupling duality in four-dimensional string theory,''
  Int.\ J.\ Mod.\ Phys.\ A {\bf 9} (1994) 3707
  [hep-th/9402002].

\bibitem{Schwarz:1994xn}
  J.~H.~Schwarz,
  ``Evidence for nonperturbative string symmetries,''
  Lett.\ Math.\ Phys.\  {\bf 34} (1995) 309
  [hep-th/9411178].


\bibitem{gsw}
M.B. Green, J.H. Schwarz and E. Witten, 1987 Superstring Theory Vols I,II, Cambridge University Press,

\bibitem{JOE}
 J. Polchinski, 1998 String Theory, Vols I,II, Cambridge University Press

\bibitem{Font:1990gx}
  A.~Font, L.~E.~Ibanez, D.~Lust and F.~Quevedo,
  ``Strong - weak coupling duality and nonperturbative effects in string theory,''
  Phys.\ Lett.\ B {\bf 249} (1990) 35.
  
   \bibitem{Eardley:1995ns}
  D.~M.~Eardley, E.~W.~Hirschmann and J.~H.~Horne,
``S duality at the black hole threshold in gravitational collapse,''
  Phys.\ Rev.\  D {\bf 52} (1995) 5397
  [arXiv:gr-qc/9505041].
  
  




 
 
 


\bibitem{Hatefi:2020jdr}
E.~Hatefi and E.~Vanzan,
``On Higher Dimensional Self-Similar Axion-Dilaton Solutions,''
Eur. Phys. J. C \textbf{80} (2020), 10
[arXiv:2005.11646 [hep-th]].

\bibitem{Goodfellow:Deep}
Goodfellow, I., Bengio, Y., \& Courville, A. ``Deep learning'', MIT press. (2016)




\bibitem{Ramsundar}
Ramsundar, B., and Zadeh, R. B. ``TensorFlow for deep learning: from linear regression to reinforcement learning," O Reilly Media, Inc., (2018).


\bibitem{Min}
Min, S., Lee, B., \& Yoon, S., ``Deep learning in bioinformatics." Briefings in bioinformatics. (2017), 851--869.

\bibitem{Choudhary:Recent}
Choudhary, K., et al. ``Recent advances and applications of deep learning methods in materials science,'' NPJ Computational Materials. (2022)


\bibitem{Collobert:unified}
Collobert, R., \& Weston, J., ``A unified architecture for natural language processing: Deep neural networks with multitask learning,'' In Proceedings of the 25-th international conference on Machine learning, (2008), 160--167. 




\bibitem{Hahnloser}
Hahnloser, et al. ``Digital selection and analogue amplification coexist in a cortex-inspired silicon circuit.'' nature  (2000): 947-951.






\end{thebibliography}
\end{document}